\documentstyle[a4,12pt]{article}
\begin{document}
\titlepage
\begin{flushright}
GPS 97/11\\
April 1997
\end{flushright}
\vskip 1cm
\begin{center}
{\bf \Large
Open supermembranes in eleven dimensions}
\end{center}

\vskip 1cm
\begin{center}
{\large Ph. Brax$^{a}$\footnote{email: brax@spht.saclay.cea.fr} $\&$ J. Mourad$^{b}$\footnote{email: mourad@qcd.th.u-psud.fr}}
\end{center}
\vskip 0.5cm
\begin{center}
$^a$ {\it Service de Physique Th\'eorique, 
CEA-Saclay\\
F-91191 Gif/Yvette Cedex, France}\\
$^b$ {\it GPS, Universit\'e de Cergy-Pontoise\\
Site Saint-Martin, F-95302 Cergy-Pontoise, France}
\end{center}
\vskip 2cm
\begin{center}
{\large Abstract}
\end{center}

\noindent
We consider open supermembranes in an eleven dimensional background. We show
that, in a flat space-time, the world-volume action is kappa-symmetric and has global space-time supersymmetry if space-time has even dimensional topological defects where the membrane can end. An example of such topological defects is  
provided by the space-time  with
boundaries  considered by Horava and Witten. In that case the world-volume action has reparametrisation anomalies whose cancellation requires the inclusion of a current algebra on the boundaries of the membrane. The role of kappa-anomalies in a general background is discussed. The tension of the membrane is related to the eleven dimensional gravitational constant 
with the aid of the Green-Schwarz mechanism allowing a consistency check of $M$-theory.
\newpage
\section{Introduction}
One of the most fascinating aspects of recent progress towards a non-perturbative understanding of string theory is the role played by an eleven dimensional theory, $M$-theory, whose low energy limit is given by the eleven dimensional supergravity.
In particular, type IIA superstring is believed to be related to $M$-theory
compactified on a circle \cite{t1,w1} and there are many indications that the $E_8\times E_8$ heterotic string is related to $M$-theory compactified
on an interval ($S^1/Z_2$) \cite{h1,h2}.
The relation of type IIA superstrings
to $M$-theory is supported by the existence 
of a closed supermembrane in eleven dimensions \cite{b1,b2} whose double dimensional reduction, that is the simultaneous reduction of both a space-time 
and a world-volume coordinate, leads to a type IIA superstring \cite{d1}. The role of
the closed supermembrane in eleven dimension was strengthened in \cite{b1}
where it was shown that kappa-symmetry of the worldvolume action leads to the 11D SUGRA equations of motion. The identification of the eleven-dimensional Kaluza-Klein states with type IIA D-0 branes \cite{w1,p1,w2} put the conjecture that the strong coupling limit of IIA superstring is eleven dimensional supergravity on a firm basis.
On the other hand, the strong coupling limit of the $E_8\times E_8$ heterotic string is believed to be given by the eleven-dimensional supergravity with one 
dimension being an interval, the $E_8$ gauge fields and their superpartners living on the boundaries of space-time \cite{h1,h2}. The arguments given by Horava and Witten in favor of this conjecture include the consideration of the degrees of freedom of the solitonic membrane and especially the cancellation of space-time gravitational anomalies \cite{h1}.
The anomaly cancellation argument was developed further in \cite{h2} (see also \cite{da,du}); together with
the requirement of local supersymmetry it allowed to determine perturbatively the Lagrangian describing  the low energy strong coupling limit of the $E_8\times E_8$ heterotic string \cite{h2}.

The aim of this paper is to examine further the membrane arguments in order to have explicit relations with the heterotic string at the world-volume level, the heterotic string being obtained by a double dimensional reduction of the open supermembrane in the same manner that type IIA superstring is obtained by a double dimensional reduction of the closed supermembrane. Open supermembranes were considered in \cite{b2} in an eleven dimensional background with no topological defects. It was noticed there that global supersymmetry cannot be achieved. We return to this issue in section 2 where we prove that global supersymmetry and kappa-symmetry can be achieved when space-time has topological defects where the membrane can end. These topological defects 
are the analogue of D-branes for type II superstrings. One of these topological defects consists of having a space-time with boundaries in the manner considered by Horava and Witten.  We also examine the boundary conditions in this section. In section 3 we
consider world-volume reparametrisation anomalies and add to the action a suitable boundary term in order to cancel anomalies. We discuss the different cases corresponding to the membrane having its two ends on one boundary of space-time or on the two boundaries. We also discuss the effect of a topological three dimensional term which 
moves the  anomaly from one boundary
to the other.
In section 4 we  
sketch the consequences of coupling the supermembrane 
to background fields, we  emphasize  the role of kappa-anomalies in deriving the classical equations of motion and especially in giving the boundary value of the four-form field strength. We show how kappa-anomalies rule out all  the configurations of the supermembrane
except the one corresponding to having one $E_8$ on each boundary of space-time.
In section 5 we show how the eleven-dimensional Green-Schwarz mechanism \cite{h2}
allows the prediction of the eleven-dimensional  gravitational constant in terms  of the membrane tension. We collect our conclusions in   section 6.
 
\section{The action and its symmetries}
We first consider a flat background. We denote the embedding of the supermembrane in eleven dimensional superspace by $Z^M(\xi^i)=(X^a(\xi^i),\theta^\alpha(\xi^i)), \ a=1,\dots 11, \alpha=1\dots 32, i=1,2,3$. The moving basis is $E^a=dX^a-i\overline{d\theta}\Gamma^a\theta,\ E^\alpha=d\theta^\alpha$.
The action of the open supermembrane, given in the Dirac-Nambu-Goto form, is
\begin{equation}
S=-T_3\left[\int_{\Sigma_3}\sqrt{-\tilde g}+\int_{\Sigma_3}\tilde C+\int_{\partial\Sigma_3} \tilde B\ \right],\label{a}
\end{equation}
where $\tilde g_{ij}$ is the induced metric 
\begin{equation}
\tilde g_{ij}=\tilde E^a_i\tilde E^b_j\eta_{ab},\ 
\end{equation}
$\tilde E^a_id\xi^i$ are the pullback of the 
local frame, $\tilde C$ is the pullback of the eleven dimensional super three-form and $\tilde B$ is the pullback
of a super two-form potential
into the world-volume \footnote{In the following we will denote the pullback into the world-volume
of a space-time form $A$ by $\tilde A$.}.
The last term in the action, which is absent for a closed supermembrane, is added in order to respect the gauge invariance:
\begin{equation}
C\rightarrow C+d\Lambda,\label{g1}
\end{equation}
where $\Lambda$ is a two-form. The action is invariant
under this transformation if the two-form potential transforms as
\begin{equation}
B\rightarrow B-\Lambda.\label{g2}
\end{equation}
Eleven-dimensional supergravity has no two-form potential in its spectrum. this is the first indication that in order to have open supermembranes one needs some topological defects localised in a supersubmanifold ${\cal M}'$ where this two-form potential can live. The boundary of the
supermembrane must lie on these topological defects.

An essential requirement for an acceptable action is its kappa-symmetry.
The kappa-transformation of the pullback of a space-time form $A$ is conveniently given by
\begin{equation}
\delta_{\kappa}\tilde A=
{\widetilde{L_{\kappa}A}},
\end{equation}
where $L_{\kappa}=d\iota_\kappa+\iota_\kappa d$ is the Lie derivative with respect to the vector field $\kappa$:
\begin{equation}
\kappa=\Delta^\alpha(Z) E_\alpha,
\end{equation}
$\Delta$ being local fermionic parameters
 constrained by
\begin{equation}
\tilde \Delta=\tilde \Gamma \tilde \Delta,\label{c}
\end{equation}
where $\tilde \Gamma$ is given by
\begin{equation}
^*\tilde\Gamma={{1}\over{3!}}
\widetilde{E^aE^bE^c\Gamma_{abc}},
\end{equation}
the $*$ represents the hodge dual on the world-volume. The kappa variation of the induced volume element is thus given by
\begin{equation}
\delta_{\kappa}\sqrt{-\tilde g}=
{{1}\over{2}}\sqrt{-\tilde g}\tilde g^{ij}\delta_{\kappa}\tilde g_{ij}=-i\epsilon^{ijk}(\gamma_{jk})_{\beta\alpha}\tilde\Delta^{\alpha}\tilde E^{\beta}_i,\label{g}
\end{equation}
where use has been made of the constraint
(\ref{c}), the definition $\gamma_id\xi^i=
\widetilde{\Gamma_aE^a}$, as well as the identity
\begin{equation}
\sqrt{-\tilde g}\tilde g^{ij}\gamma_{j}\tilde \Gamma=\epsilon^{ijk}\gamma_{jk}.
\end{equation}
Finally the kappa-variation of the action can be written as
\begin{equation}
\delta_\kappa S=-T_3\left[
\int_{\Sigma_3}\left(-i\epsilon^{ijk}(\gamma_{jk})_{\beta\alpha}\tilde\Delta^{\alpha}\tilde E^{\beta}_i +\widetilde{\iota_{\kappa}G}\right)
+\int_{\partial\Sigma_3}\widetilde{ \iota_\kappa H}\right],\label{var}
\end{equation}
where we defined the field strengths
\begin{equation}
G=dC,\ \ H=dB+C.
\end{equation}
Note that $G$ and $H$ are invariant under the gauge transformations (\ref{g1}, \ref{g2}).
The conditions for kappa-symmetry are thus
\begin{eqnarray}
G=i(\Gamma_{ab})_{\alpha \beta}E^\alpha E^\beta E^a E^b + G',\
\  \iota_{E_\alpha}G'=0\\
\iota_{E_\alpha}H=0,\ \alpha=1,\dots 32.
\end{eqnarray}
The first term  in  $G$ is closed due to a gamma matrix  identity in eleven dimensions
\cite{b1}.
The variation of the action under a global super Poincar\'e transformation generated by the vector field $P$ is given by
\begin{equation}
\delta_{P}S=-T_3\left[\int_{\Sigma_3}
\widetilde{\iota_P G} +\int_{\partial\Sigma_3} \widetilde{\iota_P H}\right].
\end{equation}
The action is thus invariant under the transformations satisfying
\begin{equation}
L_P G=0, \quad L_P H=0.
\end{equation}
The combination of kappa-symmetry and global super Poincar\'e symmetry imposes
the relations
\begin{equation}
G= i(\Gamma_{ab})_{\alpha \beta}E^\alpha E^\beta E^a E^b,\ \quad H=0.
\end{equation}
The latter relation being verified on a 
supersubmanifold, ${\cal M}'$, of the eleven dimensional super space-time  where  the boundary of the membrane lies.
The field strengths $G$ and $H$ are not independent, they are related by
\begin{equation}
dH=G|_{{\cal M}'},
\end{equation}
so we get the important constraint
\begin{equation}
G|_{{\cal M}'}=0.\label{im}
\end{equation}
Before analysing equation (\ref{im}) we 
examine the boundary conditions  
which are compatible with the requirement
\begin{equation}
\partial\Sigma_3\subset {\cal M}'.\label{em}
\end{equation}
Boundary conditions are obtained by considering an arbitrary variation of the action in order to obtain the equations of motion and demanding that the boundary terms vanish. Let the even part of ${\cal M}'$
be described by
\begin{equation}
X^{\overline {a}}=0, \ \overline{a}=11-d+1,\dots, 11,
\end{equation}
where $d$ is the dimension of ${\cal M}'$,  use the splitting $X^a=(X^{\underline{a}},X^{\overline{a}})$,
and suppose that the boundary of 
the membrane is   given by $\xi^3=0$ 
then the boundary conditions compatible with (\ref{em}) are
\begin{equation}
X^{\overline{a}}\vert_{{\partial\Sigma_3}}=0,
\quad
\tilde E_3^{\underline{a}}\vert_{\partial
\Sigma_3}=0.
\end{equation}
Supersubmanifolds ${\cal M}'$ where
(\ref{im}) holds are given by
\begin{equation}
d=2n,\quad \theta=\Gamma^1\Gamma^2\dots
\Gamma^d\theta,\ n=0,\dots 5.\label{li}
\end{equation}
These solutions preserve the supersymmetries that obey
\begin{equation}
\epsilon=\Gamma^1\Gamma^2\dots
\Gamma^d\epsilon.
\end{equation}
The case with $d=10$ represents the Horava-Witten boundary of space-time,
the case with $d=6$ is the M-theory fivebrane. The interpretation of the eleven-dimensional fivebrane as a Dirichlet
Brane for membranes was proposed previously in \cite{t2}. The possibility that membranes can end on fivebranes was pointed out in \cite{s1}
using charge conservation arguments. The existence of ninebranes (d=10) in eleven dimensions was conjectured in \cite{b3,h3,p2}. Twisted membranes with a action different from ours were considered in \cite{ald}.
The list  given in (\ref{li}) includes four new branes in eleven dimensions
(p=-1,1,3,7). The existence of these objects in eleven dimensions needs further support. Here we simply note that these objects allow  
the open supermembrane to end while preserving half of the eleven dimensional supersymmetries, whether they all exist
is beyond the scope of the present article.  
In the rest of the paper we will focus our attention on the $d=10$ case which we consider as the boundary of space-time.

\section{Reparametrisation anomalies}
Let the eleven dimensional space-time have the topology $R^{10}\times [0,l]$, with the boundaries
located at $x^{11}=0$ and $x^{11}=l$.
The boundaries of the open supermembrane must, according to the preceding section, lie
on $x^{11}=0$ and/or $x^{11}=l$. The Horava-Witten configuration, where on each boundary lives an $E_8$ supermultiplet, corresponds, as we will see,  to 
one boundary of the supermembrane on $x^{11}=0$ and the other boundary on $x^{11}=l$. The topology of the membrane
that we shall consider  is  $\Sigma_2\times I$ with $\Sigma_2$ a closed two-dimensional surface, and $I=[0,\pi]$ an interval with coordinate $\xi^3$. The case corresponding to the Horava-Witten configuration is obtained by setting $\xi^3=\pi x^{11}/l$, while the case where the two ends of the membrane lie on $x^{11}=0$ is obtained by setting, e.g., $x^{11}=l'\sin{(\xi^3)}$, with $l'<l$.

The action (\ref{a}), with these conditions,  reads, in the Polyakov form,
\begin{equation}
S=-T_3\left[{{1}\over{2}}
\int_{\Sigma_3}\Big(\sqrt{- g}(g^{ij}\tilde g_{ij}-1)
+\tilde C\Big)+\int_{\Sigma_2, \xi^3=\pi}\tilde B-\int_{\Sigma_2, \xi^3=0}\tilde B\right],
\label{aa}
\end{equation}
where we introduced the word-volume metric $g_{ij}$ as an auxiliary field.
The world-volume fields of the supermembrane can be developed in Kaluza-Klein modes in the $\xi^3$ direction. The potential $B$ is defined only on the boundary and thus has no $\xi^3$ expansion.
The resulting zero-modes coincide with 
that of the supercoordinates of the
GS formulation of the heterotic string
in ten dimensions with  $\alpha'$ given by
\begin{equation}
lT_3={{1}\over{2\pi\alpha'}},
\end{equation}
and a WZW term given by
\begin{equation}
{{1}\over{2\pi\alpha'}} \int_{\Sigma_2}\tilde B',
\end{equation}
with $B'$ a super two-form given by
\begin{equation}
B'={{1}\over{l}}\Big(B(x^{11}=l)-B(x^{11}=0)\Big)+ \iota_{E^{11}}C\label{pot}
\end{equation}
for the Horava-Witten case 
and 
\begin{equation}
2 l'T_3={{1}\over{2\pi\alpha'}},\quad B'=\iota_{E^{11}}C
\end{equation}
for the case where the two ends of the membrane are on one boundary. Note that the effective two-form (\ref{pot}) obtained by a dimensional reduction of the membrane, in the Horava-Witten configuration, involves the two different two-forms on the boundaries of space-time as well as the reduction of the three form potential.

The resulting zero-mode  action  has world-sheet reparametrisation anomalies
which were calculated in \cite{l1} and shown to be given by
\begin{equation}
X_4=dX_3=-{{1}\over{6\pi}}tr({\cal R} ^2),
\end{equation}
where $\cal R$ is the two-dimensional curvature. 
A three dimensional theory on a smooth manifold has no anomalies \cite{al}, so, as in the Horava-Witten construction, the anomalies
must be concentrated on the boundaries of the membrane. 
 One expects the anomalies
to be distributed  symmetrically between the two boundaries, that is each boundary supports half of the total anomaly
\begin{equation}
\delta\Gamma={{1}\over{2}}\int_{\Sigma_2, \xi^3=\pi}X_{2}^1+{{1}\over{2}}\int_{\Sigma_2, \xi^3=0}X_{2}^1,
\end{equation}
where $\delta X_3=dX^{1}_2$.
Note however that it is possible to add to the world-volume action the term
\begin{equation}
-
{{1}\over{2}}\int_{\Sigma_3}X_{3}\label{an} 
\end{equation}
whose effect is to remove the anomaly from one side to add it on the other side.
In fact the variation under reparametrisation of the term 
(\ref{an}) can be written as
\begin{equation}
-{{1}\over{2}}
\int_{\Sigma_3}dX_2^{1}={{1}\over{2}}
\int_{\Sigma_2, \xi^3=0}X^{1}_{2}-{{1}\over{2}}
\int_{\Sigma_2, \xi^3=\pi}X^{1}_{2}.
\end{equation}

A counter term analogous to (\ref{an}) is
 also possible in the eleven-dimensional analysis of the gravitational anomaly, namely the addition to the eleven-dimensional supergravity action of $\int_{M_{11}}I_{11}$, where $I_{12}=dI_{11}$ represents the ten dimensional gravitational anomaly. The  $E_8\times E_8$ heterotic string, with each $E_8$ propagating on one boundary, was related in \cite{h1,h2} to the case were the anomaly is symmetrically distributed between the two boundaries of space-time. The inclusion of the topological term $\int I_{11}$ allows to argue for the possibility of having  $E_8\times E_8$ or a $SO(32)$ on one boundary and no matter on the other boundary. This picture is compatible, as we will see, with the world-volume reparametrisation anomaly cancellation but needs further analysis
on the eleven-dimensional level where 
the Green-Schwarz mechanism 
is not expected to arise in a natural way from the Chern-Simons terms of eleven-dimensional supergravity as is the case for the Horava-Witten case. We shall return to this point in  the next sections. 

\begin{figure}
\vspace{7 cm}
\special{hscale=60 vscale=60 voffset=0 hoffset=10 psfile=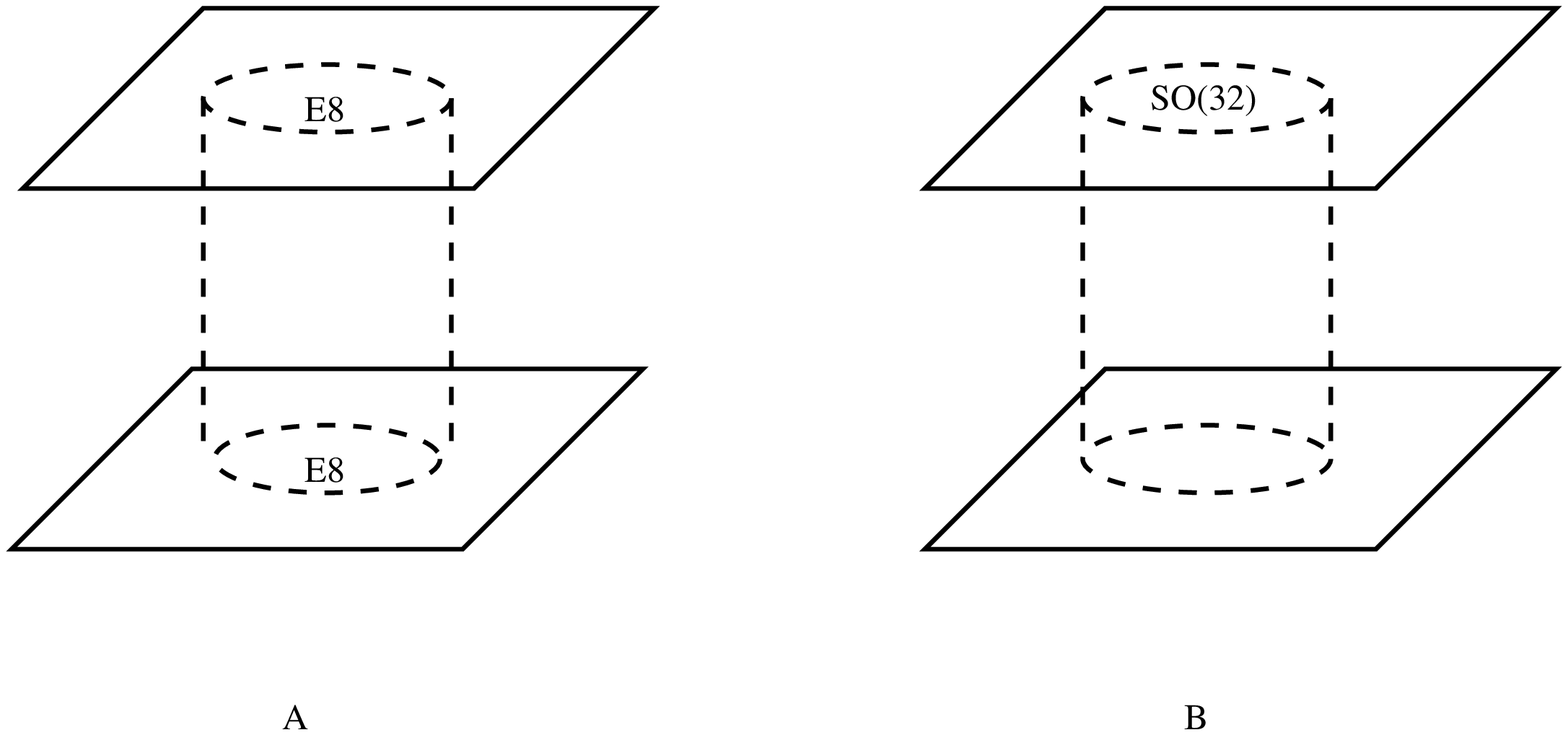}
\end{figure}
\begin{figure}
\vspace{7 cm}
\special{hscale=60 vscale=60 voffset=0 hoffset=10 psfile=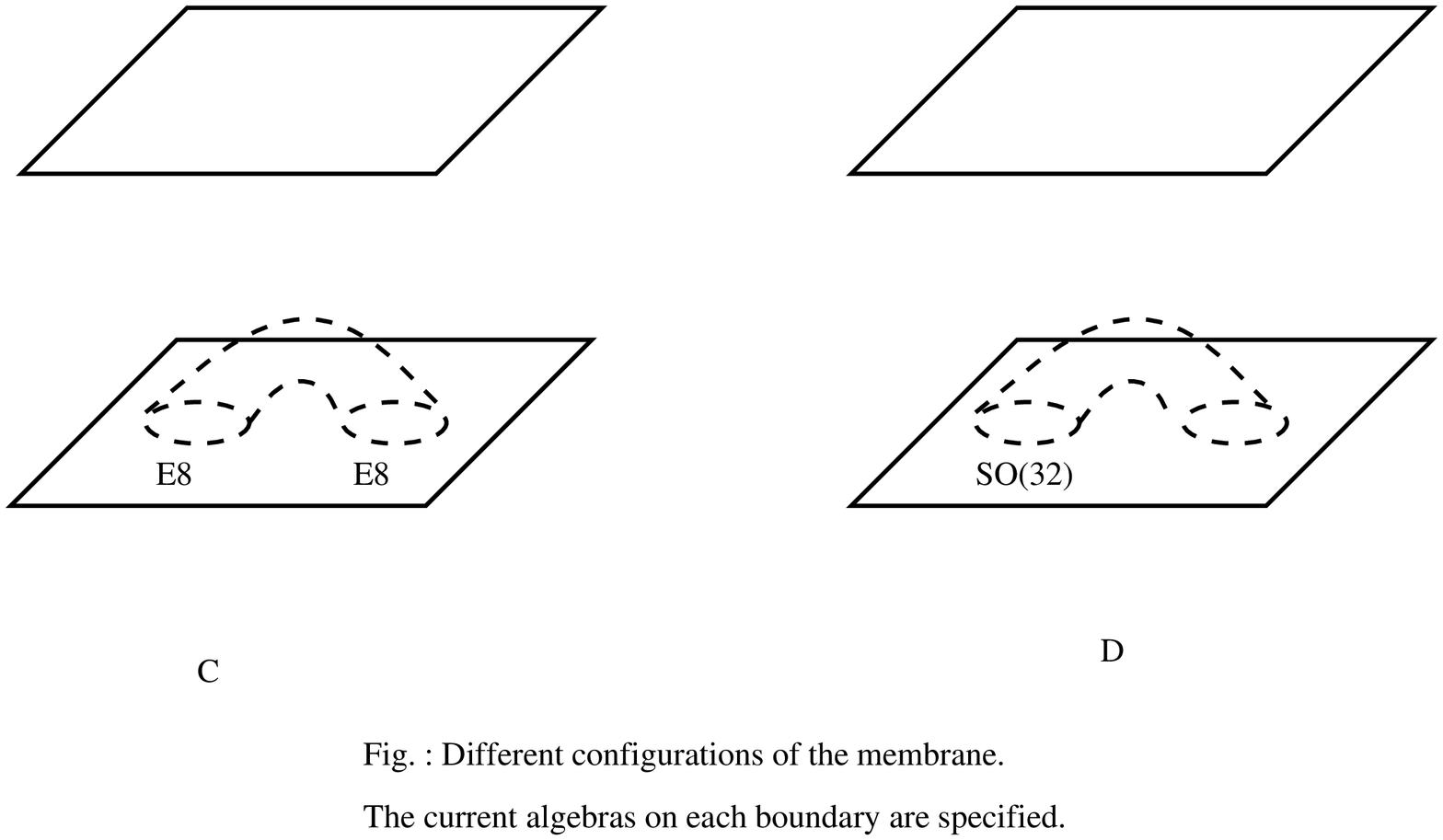}
\end{figure}

Suppose first that the anomaly on the world-volume is distributed symmetrically between the two boundaries, then one has to
add 16 two-dimensional Majorana-Weyl fermions on each boundary. The 16 fermions form an $E_8$ current algebra.  We can have two situations: i) the two boundaries of the membrane are on $x^{11}=0$, then one has 
an $E_8\times E_8$ gauge multiplet on $x^{11}=0$ (fig. C); ii) one boundary of the membrane is on $x^{11}=0$ and the other 
is on $x^{11}=l$, then one has one $E_8$ propagating on each boundary and recovers the Horava-Witten construction (fig. A).

Suppose next that, with the aid of the three dimensional topological term, all the anomaly is concentrated on one boundary, then one has to add on this boundary 32 Majorana-Weyl fermions. They form the 
current algebra of $SO(32)$ or $E_8\times E_8$. This case, wherever are  the two boundaries of the membrane (fig.B, D), corresponds to a $SO(32)$ or $E_8\times E_8$ multiplet propagating on one boundary of space-time.

The boundary term required for the anomaly cancellation which must be added to the action (\ref{aa}) (and depending on the situation considered to (\ref{an})) is of the form
\begin{equation}
{{1}\over{4\pi\alpha'}}\int_{\Sigma_2}
\sqrt{-g}\psi^{t}\partial_-\psi,
\end{equation} 
where $g$ is the restriction of the three dimensional metric on the boundary and $\psi$ represents a multiplet of $16$ or $32$ world-sheet Majorana-Weyl fermions.  
These fermions,  in flat space-time, must be invariant under kappa transformations and global supersymmetry. 

In brief, we found in this section 
that the cancellation of world-volume repara\-me\-trisation anomalies leads to
five different configurations depending
on whether the membrane starts and ends  on the same boundary and whether the current algebra is on either the same boundary or on both. In the next section all of these configurations except the Horava-Witten case will be ruled out by the requirement of one-loop kappa-anomaly cancellation in a general background.
 
\section{General background and kappa anomalies}

The generalisation of the preceding action to curved space-time with a coupling to a super Yang-Mills potential $A$ on the boundary is straightforward.
The bulk action is analogous to the closed membrane action considered in \cite{b1} and  the fermion boundary term is analogous to the one existing in the heterotic string \cite{b4}.
The classical kappa-symmetry of the action 
imposes constraints on the background superfields which are, after field redefinitions, those of the superfield 
formulation of eleven-dimensional supergravity \cite{b1} and supersymmetric Yang-Mills 
in ten dimensions \cite{b4} with the boundary condition
\begin{equation}
dH=G\vert_{{\cal M}'}=0.\label{bo}
\end{equation}
As explained in \cite{t3} 
for the ten dimensional heterotic string,
the coupling between the two sectors 
arises when one considers quantum kappa-anomalies. As noted in the preceding section, the zero-modes of the action is the GS formulation of the heterotic string
so the action is kappa-anomalous with the anomalies concentrated on the boundaries of the membrane. The one loop total anomaly  found in \cite{c1} may be written as
\begin{equation}
\delta_{\kappa}\Gamma=-{{1}\over{8\pi}}\int_{\Sigma_2}\widetilde{\iota_\kappa\omega},
\end{equation}
with the the three- superform $\omega$ given by
\begin{equation}
\omega=\omega_{3YM}-\omega_{3L},
\end{equation}
$\omega_{3YM}$ and $\omega_{3L}$ being
the Yang-Mills and Lorentz Chern-Simons forms, so that
\begin{equation}
d\omega=tr(F^2)-tr(R^2).
\end{equation}

Let us first consider the case where 
one boundary of the membrane is on $x^{11}=0$ and the other on $x^{11}=l$
and where there  is on each boundary 
 an $E_8$ current algebra (fig. A). This case is characterized, as we have seen, by a symmetrical distribution of the world-sheet reparametrisation and space-time gravitational anomalies.
The  kappa-anomaly in this case reads
\begin{equation}
\delta_{\kappa}\Gamma=-{{1}\over{8\pi}}
\left[\int_{\Sigma_2, \xi^3=\pi}\widetilde{\iota_\kappa\omega_1}
+\int_{\Sigma_2, \xi^3=0}\widetilde{\iota_\kappa\omega_2
}\right],
\end{equation}
with $\omega_i=\omega_{3YM\ i}-{{1}\over{2}}\omega_{3L}$ the index $i$ representing the two different $E_8$ gauge fields.
The structure of these anomalies is similar to the last term in (\ref{var}). This suggests that their cancellation can be achieved by the replacement of $H$ by $H'$ with 
\begin{eqnarray}
H'= H+{{1}\over{8\pi T_3}}\omega_1,\quad at\ x^{11}=l\\
H'= H-{{1}\over{8\pi T_3}}\omega_2,\quad at\ x^{11}=0.
\end{eqnarray}
Using the fact the $dH'=0$
we get 
\begin{equation}
G\vert_{{\cal M}', x^{11}=l}={{1}\over{8\pi T_3}}
\Big(tr(F^2_1)-{{1}\over{2}}tr(R^2)\Big)\equiv -{{1}\over{8\pi T_3}} I_4
\label{bou}
\end{equation}
and an analogous result with the opposite sign at $x^{11}=0$ which replaces the boundary condition (\ref{bo}). This constitutes an alternative derivation of the result of Horava and Witten where the value of $G$ at the boundary was fixed by the requirement of local supersymmetry and gravitational anomaly cancellation.
Our relation is expressed in terms of the membrane tension whereas the relation
of \cite{h2} is expressed in terms of the gauge and coupling constant.  This will allow, in the next section, a consistency check of $M$ theory.   

The other configurations can be studied in a similar way. The above mechanism for the cancellation of  kappa-anomalies 
is not possible when the membrane starts and ends on the same boundary (figs. C,D)  because the 
the $H$ cannot de modified in two different ways on the same space-time boundary.

Consider next 
the $SO(32)$ (or $E_8\times E_8$) case where the gauge supermultiplet is on $x^{11}=l$ and 
the boundary of the membrane carrying the $SO(32)$ (or $E_8\times E_8$) current algebra is on $x^{11}=l$ (fig. B).
The  one loop kappa-anomaly in this case is given by
\begin{equation}
\delta_{\kappa}\Gamma=-{{1}\over{8\pi}}
\left[\int_{\Sigma_2, \xi^3=\pi}\widetilde{\iota_\kappa\omega'}+\int_{\Sigma_2, \xi^3=0}\widetilde{\iota_\kappa\omega''}
\right]+{{1}\over{2}}\int_{\Sigma_3}\delta_{\kappa}X_3,
\end{equation}
where $\omega'=\omega_{3YM}-{{1}\over{2}}\omega_{3L}$ and $\omega''=-{{1}\over{2}}\omega_{3L}$.
The fact that all the anomaly originating from the gauge background is located at $x^{11}=l$ is due to the fact that the gauge part of the anomaly is due to the heterotic fermions. The precise value of $\delta_\kappa X_3\neq 0$ is not relevant.
The fact that the anomaly mixes 
terms which are pullback fields and world-volume fields makes it impossible to modify $H$ in such a way as to cancel the anomaly.
So we conclude that this case is also ruled out since it is kappa-anomalous.

In brief, only the configuration cooresponding to the Horava-Witten case (fig.(A)) is non-anomalous.  At the eleven-dimensional level, the same result has an apparently different origin. In fact 
the anomaly cancellation in eleven dimensions is due to the presence in the Lagrangian of two topological terms, the first is  the Chern-Simons interaction 
$\int CG^{2}$ and the second is of the form 
$\int C X_8$ with $X_8$ a quartic polynomial in 
the curvature given  by
\begin{equation}
X_8=-{{1}\over{8}}tr(R^4)+{{1}\over{32}}\left(tr(R^2)\right)^2;
\end{equation}
the latter term can de deduced from anomaly cancellation argument in the five-brane \cite{duf,w4} or from the type IIA superstring after a compactification on $S^1$ \cite{cw}. For the Green-Schwarz mechanism to work, with no other terms added to the Lagrangian, the anomaly, on each boundary, must be factorizable as
$I_{12}\propto I_4(I_4^2/4-X_8)$ where $I_4$ is proportional to the value of $G$ at the boundary. It turns out that  this
is only true for   
the Horava-Witten configuration. 

\section{The Green-Schwarz mechanism
in eleven dimensions}

In the preceding section it was shown that $H$ must be modified in order to cancel 
kappa-anomalies as
\begin{equation}
H'=C+dB+{{1}\over{8\pi T_3}}\omega_1
\end{equation}
on $x^{11}=l$. Under a gauge or Lorentz transformation $H'$ must be invariant, this fixes the variation of $C+dB$ as
\begin{equation}
\delta(C+dB)=- {{1}\over{8\pi T_3}}d\omega_2^1,
\end{equation}
where $\delta\omega_1=d\omega_2^{1}$.
By a gauge transformation (\ref{g1}, \ref{g2}) it is possible to choose
\begin{equation}
\delta C=0,\quad\delta B=-{{1}\over{8\pi  T_3}}\omega_2^1.\label{tra}
\end{equation}
The resulting  transformation of the effective two-form  potential $B'$ (\ref{pot}) is thus given by
\begin{equation}
\delta B'=-{{\alpha'}\over{4}}
(\omega^1_{2YM1}+
\omega^1_{2YM2}-\omega^1_{2L}),
\end{equation}
which is the usual transformation of the two-form potential in ten dimensional string theory.
The transformations  (\ref{tra}) raise the problem of understanding how the Green-Schwarz mechanism can cancel the anomalies in eleven dimensions since $C$ is invariant.
The solution to this problem consists in first noticing that the topological terms
in the low energy limit of $M$-theory \footnote{Our conventions are those of the second reference in {\cite{da}} they are related to those of \cite{h2} by $C=6\sqrt{2}C^{HW},G=\sqrt{2}G^{HW}$ and to those of \cite{duf} by $k_{11}^2=2k_{11}^{2\ DLM}$.},
\begin{equation}
S_{T}=-{{1}\over{6k_{11}^2}}\int_{M_{11}}C\wedge G^{2}+{{T_3}\over{12(2\pi)^{4}}}
\int_{M_{11}}C\wedge X_8,
\end{equation}
 are not invariant under the gauge tranformation (\ref{g1}, \ref{g2}) \cite{du}; so in order to restore this gauge invariance one has to add to the action, the boundary terms
\begin{equation}
\Delta S_{T}=-{{1}\over{6k_{11}^2}}\int_{\partial M_{11}}B\wedge G^{2}+{{T_3}\over{12(2\pi)^{4}}}
\int_{\partial M_{11}}B\wedge X_8,
\end{equation}
The gauge where (\ref{tra}) is true is natural because the anomaly is concentrated at the boundary and its cancellation is realised by fields which live on the boundary. The gauge chosen in \cite{h2, da} corresponds to setting $\delta B=0$ and this requires
to have a variation of $C$ which is non-zero in the bulk \cite{du}.
Recall that the anomaly at $x^{11}=l$ is given by
\begin{equation}
\delta\Gamma=-{{1}\over{48(2\pi)^5}}\int
\omega^{1}_{2}\wedge \Big({{I_4^2}\over{4}}-X_8\Big).\label{ano}
\end{equation}
On the other hand, the variation of the ten dimensional topological terms is given by
\begin{equation}
 \delta\Delta S_{T}={{1}\over{6 k_{11}^2}}\left({{1}\over{8\pi T_3}}\right)^3
\int \omega_2^1\wedge I_4^2-{{1}\over{48(2\pi)^{5}}}
\int \omega_2^1\wedge X_8.
\end{equation}
Comparaison with (\ref{ano}) shows that the
$X_8$ part of the anomaly cancels
 and that the remaing part is cancelled  if the membrane tension and the gravitational constant are related by
\begin{equation}
k_{11}^2={{2\pi^2}\over{T_3^3}}.
\label{rel}
\end{equation}
A relation similar to (\ref{rel}) was derived in different manners: in \cite{duf} (see also \cite{da, w3}) it was derived  
with the aid of the quantization of the flux of $G$, in \cite{da} it was derived using the D-brane tension formulae given in
\cite{p1}. Here we have given yet another derivation combining the world-volume and space-time anomaly cancellation. 
This is an additional indication of the important role that open supermembranes 
have in $M$-theory.

Before ending this section let us note that 
the gauge coupling constant can  also be determined in this framework. In fact the
value of $G$ at the boundary determined in 
\cite{h2} reads, in our notations,
\begin{equation}
G\vert_{x^{11}=l}=-{{k_{11}^2}\over{2\lambda^2}} I_4,
\end{equation}
where $\lambda$ is the gauge coupling constant.
Comparing this relation with our relation (\ref{bou}), we get
\begin{equation}
\lambda^2=4\pi T_3 k_{11}^2=8{{\pi^3}\over{T_3^2}}=2\pi(4\pi k_{11}^2)^{2/3}.
\end{equation}
The last expression is the one derived in \cite{h2}.

\section{Conclusion}
We showed that  open supermembranes  
propagate in eleven dimensions provided the boundaries of the membrane  lie on an even dimensional submanifold, the action having half the supersymmetries of eleven-dimensions. We studied  
the open membrane in the space-time considered by Horava and Witten. The cancellation of world-volume reparametrisation anomalies
allows many configurations depending on 
whether the membrane starts and ends on either the same space-time boundary or on both; and on how to distribute the anomaly between the two boundaries of the membrane. We showed that all these configurations, except the case where each boundary carries an $E_8$ current algebra and lies on a different space-time boundary, are kappa-anomalous. This provides a further confirmation, at the supermembrane level, of the result of Horava and Witten that the only possibility on $M_{10}\times S^1/Z_2$ is $E_8\times E_8$ with each factor propagating on a boundary.
In addition it gives an alternative derivation of the value at the boundary of the four-form field strength and the relation between the membrane tension and the gravitational constant. This relation is compatible with other independent derivations and thus constitutes a positive
consistency check of $M$-theory.

A similar study for membranes in an eleven-dimensional background with five-branes where the membrane can have its boundaries would be of interest, especially that the five-brane action is now known
\cite{ba,ag}. 

{\it Acknowledgements}: We would like to thank
E. Dudas for many helpful remarks.

\end{document}